\documentstyle[12pt,epsfig]{article}
\def \ins#1#2#3#4#5#6 {
  \begin{figure}[#1]
    \begin{center}
      \psfig{file=#2,width=#3,height=#4,angle=0}
      \caption{#5}
      \label{#6}
    \end{center}
  \end{figure}
  }
\begin{document}

\begin{center}
{\Large \bf
Cross Sections of Various Processes in Pbar~P-Interactions}
\end{center}

\begin{center}
V.V. Uzhinsky, A.S. Galoyan\\
Joint Institute for Nuclear Research\\
Dubna, Russia
\end{center}

\begin{minipage}{15cm}
Problem of description of total cross sections of
$\bar pp$- and $pp$-interactions is considered within the framework of
the Regge theory. Parameters of the pomeron exchange, ordinary meson
exchange and exotic meson exchange with hidden baryon number are
determined at fitting of experimental data assuming the identical
interaction radii of the reggeons with nucleon. An expression is
proposed for cross sections of various reactions in the
$\bar pp$-interactions - cross sections of one and two
string creations needed for Monte Carlo simulation of the reactions,
and the cross sections are calculated.
\end{minipage}

\section{Introduction}
Baryon number flow over large rapidity intervals attracts much
theoretical attention now \cite{Kharzeev, Kopeliovich, Guylassy}. It
is assumed that observed increase of baryon multiplicity over
antibaryon multiplicity in the central rapidity region of
nucleus-nucleus interactions \cite{NA35, NA44, NA49, RHIC} is
connected with a specific space configuration of the gluon field in
baryons. The configuration of local maxima of the energy density in
the baryon can have the Mercedes star form
\cite{SJ-annig,QCD-baryon,Bali,Simonov} (see left part of Fig. 1). The
global maximum is associated with a string junction in the string
model of baryons \cite{Baryon-strings}. It is possible, of course, a
$\Delta$-form of baryons (see consideration in Refs.
\cite{Baryon-strings,SJ-annig,Bali,Simonov}). It is assumed very often
that baryon has a compact diquark (see right part of Fig. 1).
\begin{figure}[cbth]
\begin{center}
\psfig{file=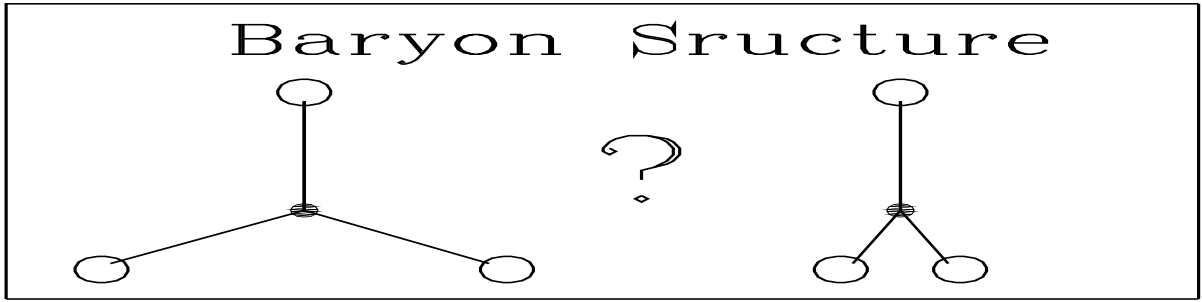,width=160mm,height=40mm,angle=0}
\caption{A possible baryon structure.}
\end{center}
\end{figure}

Depending on the assumed structure there can be various final states
in baryon-baryon interaction, especially in $\bar pp$ ones. Some of
them are shown in Fig. 2. In the process of Fig. 2a the string
junctions from colliding baryons represented by dashed lines
annihilate, and three strings are created.
\begin{figure}[cbth]
\begin{center}
\psfig{file=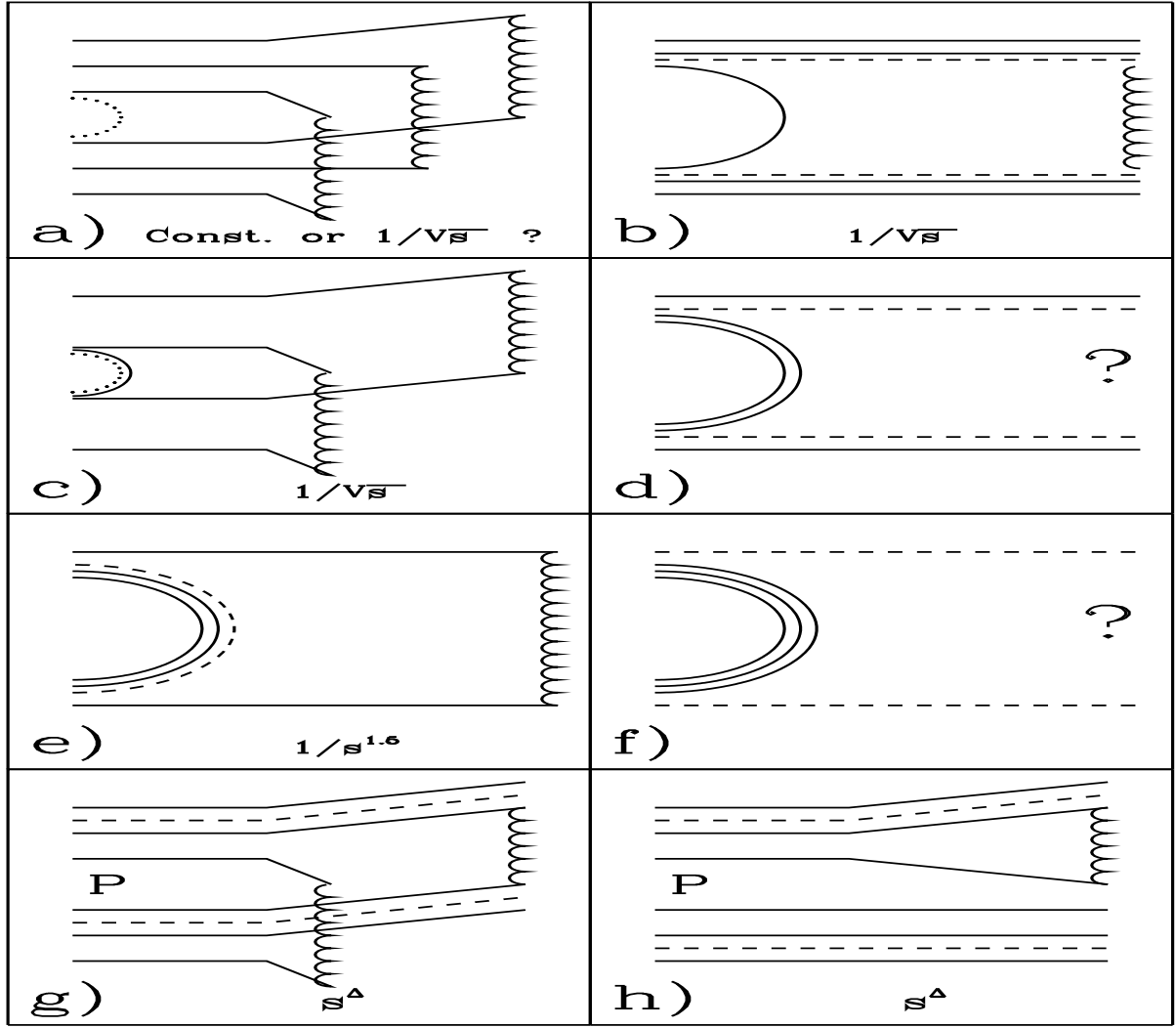,width=160mm,height=140mm,angle=0}
\caption{Possible processes in $\bar pp$-interactions}
\end{center}
\end{figure}
In the process of Fig. 2b quark and antiquark annihilate, and one
string appears. In the process of Fig. 2c quark, antiquark and the
string junctions annihilate. In the process of Fig. 2d diquarks
annihilate, and so on. Additional to the annihilations there can be
creation of two strings (see Fig. 2g) or creation of one string (Fig.
2h) at the diffraction dissociation of baryon. The process 2f can
be responsible for glueball production, and the process 2d -- for
exotic meson production. So, various final states included glueball
and exotic meson production can be in $\bar pp$ interactions. One can
expect that creation of the antiproton factory at GSI
\cite{GSI-projects} will allow to study most of them.

It is very important for future experimental investigations to estimate
production cross sections and dominating modes of the exotic state
decays. In the paper estimations of various reaction cross sections obtained
within the framework of the standard Regge phenomenology are
presented. They can be useful at determination of background condition
of the experiments.

According to the Regge phenomenology cross sections of various
reactions in the $\bar pp$ interactions are connected with various
cuts of the elastic scattering amplitude. In
the simplest approach the amplitude is determined as a sum of
contributions of different diagrams with different reggeon exchanges,
with vacuum (pomeron -- P) and non-vacuum ones ($f$, $\omega$, $A_2$,
...).
\begin{equation}
F_{\bar pp}= F_P + F_f + F_{\omega} + \ldots
\end{equation}
\begin{figure}[cbth]
\begin{center}
\psfig{file=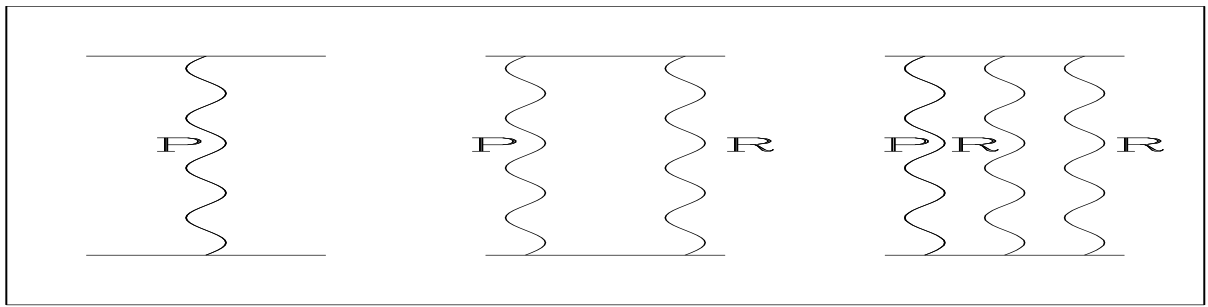,width=160mm,height=40mm,angle=0}
\caption{Reggeon exchanges in the elastic scattering.}
\end{center}
\end{figure}

Contribution of a diagram with one reggeon exchange in the impact parameter
representation has a form\footnote{$^)$ Basis of the Regge theory see
in Ref. \protect{\cite{Collins}}.}$^)$;
\begin{equation}
F_{R}(\vec{b})= i \frac{\eta _{R} g_{\bar pR}(0) g_{pR}(0)}
{R^{2}_{\bar pR}+R^{2}_{pR}+ \alpha'_{R}\xi}
e^{(\alpha_{R}(0)-1)\xi}
e^{-\vec{b}^{2}/4(R^{2}_{\bar pR}+R^{2}_{pR}+\alpha'_{R} \xi)}.
\label{eq1}
\end{equation}

The quantities $\alpha _{R}(0)$ and $\alpha ^{'}_{R}$ are the
intercept and slope of Regge trajectory $R$, $g_{pR}$ and
$g_{\bar pR}$ are constants of the reggeon interaction with the proton and
antiproton, respectively. Parametrization of the interaction vertex
dependence on 4 momentum transfer is the gaussian one,
$g_{R}(t)=g_{R}(0)e^{R^{2}_{R}t}$. $R_R$ is the effective radius of the
reggeon interaction with the nucleon. $\eta _{R}$ is the signature
factor, $\eta _{R}=1+i~ctg(\pi \alpha_R/2)$ for pole with positive
signature, and $\eta _{R}=-1+i~tg(\pi \alpha _{R}/2)$ for pole with
negative signature. $\xi =\ln(s/s_0)$, $s$ is the square of
CMS energy of the $\bar pp$ interaction, $s_0=1$ GeV$^2$.

The following parameter values are usually used for the pomeron
exchange:
$$
\alpha_P(0)=1.08, ~~~ \alpha '_P=0.13
$$
$$
\alpha_P(0)=1.12, ~~~ \alpha '_P=0.13
$$

Increase of the yield of the one-pomeron exchange with energy growth
violates the unitarity condition ($Im F(\vec b)\leq 1$), and the
problem of taking into account contributions of the more complicated diagrams
arouses. One of its solution is the eikonal approach\footnote{$^)$
Expression (\ref{eq1}) is obtained in the assumption
\protect{\cite{TM88}} that the nucleon does not convert into a jet of
particles with low mass.}$^)$
\begin{equation}
F(\vec b)=i \left\{ 1~-~exp\left[ i \sum_R F_R(\vec b)\right]\right\}.
\label{eq2}
\end{equation}

This solves also the problem of the unitarization at low energies
where increase of the contributions of the non-vacuum exchanges takes
place (see Refs. \cite{E2-86-471,Echaya}). Typical values of $\alpha_M$
of the meson exchanges are  $\simeq 0.5$, the nucleon ones -- $\alpha_N \simeq
-0.5$, and so on. However a task of a correct definition of
inelastic process cross sections appears.

The total cross section in the chosen representation has a form:
\begin{equation}
\sigma^{tot}= 2Im \int d^2b \left\{ 1~-~exp\left[
i \sum_R F_R(\vec b)\right]\right\}.
\label{eq3}
\end{equation}

The total elastic cross section is given by the expression:
\begin{equation}
\sigma^{el}= \int d^2b \left\{ 1~-~
2Re~exp\left[ i \sum_R F_R(\vec b)\right]~ +~
exp\left[ -2 Im \sum_R F_R(\vec b)\right]\right\}.
\label{eq4}
\end{equation}

The inelastic cross section is
\begin{equation}
\sigma^{in}= \int d^2b \left\{ 1~-~
exp\left[ -2 Im \sum_R F_R(\vec b)\right]\right\}.
\label{eq5}
\end{equation}

Considering only the pomeron exchanges, in the high energy
approximation, $\sigma^{in}$ can formally be represented as
\begin{equation}
\sigma^{in}=\sum_{n=1}^{\infty }\sigma_n=
\sum_{n=1}^{\infty } \int d^2b \frac{[2 Im F_p(\vec b)]^n}{n!}
exp\left[ -2 Im \sum_R F_R(\vec b)\right]
\label{eq6}
\end{equation}

The same result can be obtained at an application of the asymptotic
Abramovski-Gribov-Kancheli cutting rules \cite{AGK}, AGK rules.
At this, one can find that $\sigma_n$ is the cross section with
$n$ cut pomerons. According to the existing interpretation
\cite{DPM} $\sigma_n$ is the cross section of the process with
the creation  of $2(n-1)$ quark--antiquark strings and 2
quark-diquark strings. If we assume that each string produces at
least one meson, we obtain a restriction on $n$,
$$
n \leq n_{max}=\sqrt{s}/2M_M,~~~ M_M\simeq 0.5~~GeV.
$$
Therefore the series (\ref{eq6}) must be effectively restricted, or a
principle of the cut diagram yield re-summation must be formulated.
It is disirable to save the simple expressions for $\sigma^{tot}$ and
$\sigma^{el}$.

The simplest recipe is,
\begin{equation}
\sigma^{in}= \sum_{n=1}^{n_{max}} \int d^2b~ e^{ -2 Im F_p}
\left( e^{ 2 Im F_p/n_{max}}-1\right)\times
e^{ \frac{n-1}{n_{max}} 2 Im F_p}.
\label{eq7}
\end{equation}

The other simple prescription approbated by the additive quark model  is,
\begin{equation}
\sigma^{in}= \sum_{n=1}^{n_{max}} \int d^2b~
e^{ -\frac{n_{max}-n}{n_{max}} 2 Im F_p}
\left[1- e^{ -2 Im F_p/n_{max}}\right]^n.
\label{eq8}
\end{equation}

The two different representations lead to two different
interpretations of the AGK rule application at finite energies, and
two different scenarios (mechanisms) of the interaction can be
created, respectively.

At a consideration of the other reggeon exchanges we have additional problem
of selection of cross sections connected with cutting of different
reggeons. For example, as known jet production takes place in
nucleon-nucleon interactions at super high energies. To take them into
account the phase function, $\chi =\sum_R F_R$, is represented as
$\chi =\chi_{soft}+\chi_{jet}$. This gives
\begin{equation}
\sigma^{in}= \int d^2b~ \left( 1-
e^{-2 Im (\chi_{soft}+\chi_{jet})}\right)=
\sum_{m,n=0,~m +n\neq 0}^{\infty } \sigma_{m,n},
\label{eq10}
\end{equation}
$$
\sigma_{m,n}= \int d^2b~ \frac{[2 Im \chi_{soft}]^m}{m!}
\frac{[2 Im \chi_{jet}]^n}{n!} e^{-2 Im (\chi_{soft}+\chi_{jet})}.
$$
$\sigma_{m,n}$ is interpreted as a cross section of a process with
creation of $2n$ jets and $2m$ strings.

In the case of $\bar pp$ interactions at low energies one needs to
consider both energy restriction and different reggeons.

\section{General expressions}
Most of the existing applications of the Regge phenomenology to the
description of the $\bar pp$ interactions consider only one-reggeon
exchanges,
\begin{equation}
F(\vec b)=\sum_{R}F_R=F_p + F_f + F_{\omega} + ...
\label{eq11}
\end{equation}
In this case the imaginary part of the amplitude is a sum of positive
defined terms.

Performing the unitarization and finding the inelastic cross section,
we have
\begin{equation}
\sigma^{in}=\int d^2b~ \left[1 - e^{-2 Im F_p - 2 Im F_f -
2 Im F_{\omega}-...}\right].
\label{eq12}
\end{equation}

Of course, one can use for expansion of the $\sigma^{in}$ an analogy of
the expression (\ref{eq10}) and try to take into account the
finiteness of the energy. At this one can meet a problem of physical
interpretation. The matter is that the expression (\ref{eq6})
was obtained in the assumption of non-planar
vertexes of the pomeron interactions with the nucleon. At low energies
contributions of planar diagrams are dominated in the amplitude.

Consideration of the planar diagrams at the calculation of the $\sigma_n$
is a rather difficult task. Though, assuming identical dependencies of
all $F_R$ on $b$ and the same longitudinal structure of the
reggeon-baryon interactions we propose the following expression,
\begin{equation}
\sigma^{in}=\sum_i \sigma_i,~~~
\sigma_i=\frac{ Im F_i(0)}{\sum_R Im F_R(0)}
\int d^2b~ \left[1 - e^{-2 Im \sum_R F_R}\right].
\label{eq13}
\end{equation}
Each quantity $\sigma_i$ is the sum of contributions of diagrams shown
in Fig. 4 in the assumption that states $X$ associated with cuts of
different reggeons are ortogonal to each other.
\begin{figure}[cbth]
\begin{center}
\psfig{file=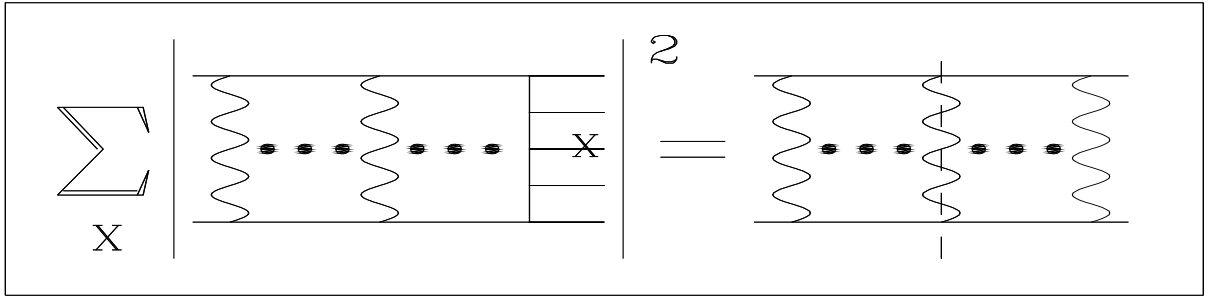,width=160mm,height=40mm,angle=0}
\caption{Graphs contributed into the inelastic cross section.}
\end{center}
\end{figure}

It is obvious that there must be gluon exchanges in the $\bar pp$
interactions. Therefore the pomeron contributions must be in the
elastic scattering amplitude, and a term proportional to
$s^{\alpha_p(0)-1}$ must be in the phase function

One meson exchange must be responsible for the quark and antiquark
annihilation. Therefore a corresponding term proportional to
$s^{\alpha_M(0)-1}$ must be in the phase function. Because the typical
values of $\alpha_M(0)\simeq 0.5$, the corresponding term is
proportional to  $1/\sqrt{s}$.

The baryon annihilation is connected with the processes shown in
Figs. 2a, 2c, 2e \cite{SJ-annig}. According to the estimations of
Ref. \cite{Eylon} the corresponding reggeon trajectories can have
intercepts \footnote{ It was used the expression (4) of Ref.
\protect\cite{Eylon} and $\alpha_p=1.$}$^)$
\begin{eqnarray}
\alpha_a(0)=2 \alpha_B(0) - 1 + ~ (1-\alpha_M(0)),  \\
\alpha_c(0)=2 \alpha_B(0) - 1 + 2 (1-\alpha_M(0)),  \\
\alpha_e(0)=2 \alpha_B(0) - 1 + 3 (1-\alpha_M(0)),
\end{eqnarray}
where $\alpha_B$ is the intercept of the leading baryon trajectory.

Using as in Ref. \cite{Eylon,SJ-annig} the values of the
parameters $\alpha_B(0)\simeq 0$, $\alpha_M(0)=0.5$, we have
$\alpha_a=0.5$, $\alpha_c=0$ and $\alpha_e=-0.5$. Therefore terms
proportional to $1/\sqrt{s}$, $1/s$ и  $1/s^{1.5}$ can be in the phase
function.

The choice of the effective
intercept $\alpha_B(0)\simeq 0$ is seemed reasonable \cite{Webber}
because $\alpha_N(0)=-0.35$ and
$\alpha_\Delta(0)=+0.15$ (see  Ref. \cite{Collins}). Though as it was
noted in Ref. \cite{E2-86-471,Echaya} it is not known what kind of the
exchanges are dominated, and a pessimistic choice
$\alpha_B(0)\simeq \alpha_N(0)=-0.35$ gives the values $\alpha_a
\simeq -0.2$, $\alpha_c \simeq -0.7$ and $\alpha_e \simeq -1.2$.
The value of $\alpha_N(0)=-0.5$ is often used last years, it leads to
$\alpha_a \simeq -0.5$, $\alpha_c \simeq -1$ and $\alpha_e
\simeq -1.5$.

In order do not miss any possibilities we have parametrized the phase
function by the following form,
\begin{equation}
\chi (0)_{\bar pp}=a_1 \eta_P S^\Delta ~+~ \eta_M \frac{a_2}{S^{a_3}}
~+~\eta_c\frac{a_4}{S^{1}} ~+~\eta_e\frac{a_5}{S^{1.5}}
~+~\eta_{c'}\frac{a_6}{S^{2}} ~+~\eta_{e'}\frac{a_7}{S^{2.5}}
\end{equation}

A choice of the signature factors is a rather complicated question. It
is known \cite{Collins} that the meson trajectories have a strong
exchange degeneracy. Therefore  in the first approximation we
assume $\eta_M=1$ \cite{E2-86-471}. We suppose a negative signature of
the pole responsible for the process of Fig. 2c. We will try to
estimate the signature of the pole connected with the process of
Fig. 2e. Criteria of the rightness of the choice can be a successful
description of the total cross sections of the $\bar pp$ and $pp$
interactions, and the ratios of the real to imaginary part of the
corresponding elastic scattering amplitudes. In accordance with saying
above we choose the phase function of the $pp$ scattering in the form,
\begin{equation}
\chi (0)_{pp}=a_1 \eta_P S^\Delta ~+~ \frac{a_2}{S^{a_3}}
~-~\frac{a_4}{S^{1}} ~\pm ~\frac{a_5}{S^{1.5}}
~-~\eta_{c'}\frac{a_6}{S^{2}} ~-\eta_{e'}\frac{a_7}{S^{2.5}}
\end{equation}

The dependence of the phase function
was determined as in Ref. \cite{E2-86-471},
\begin{equation}
\chi (\vec b)=\chi (0) e^{-b^2/2B_{eff}},
\end{equation}
$$
B_{eff}=B_0 + 2 \alpha_p' \xi,~~~B_0=10~GeV^{-2},~~~
\alpha_p'=0.13~GeV^{-2}.
$$

\section{Fitting procedure}
We have used at fitting of the parameters the experimental data on the
total $\bar pp$ and $pp$ interaction cross sections collected in the
Particle Data Group data base \cite{PDG}. The fit is a
rather complicated procedure because:
\begin{enumerate}
\item The number of the parameters is large, and the fitting results
are unstable;

\item The signatures of the poles are not determined quite well;

\item Application region of the considered approach is not known.

\end{enumerate}

At the beginning we performed the fit taking only the two first
terms of the phase function. Such oversimplified approach can be
useful at a rough estimation of the cross sections. We present results
of the fit in table. 1 and description of the experimental data at
$p_{lab}\geq 5$ GeV/c in Fig.  5 \footnote{ Behavior of the
approximating functions below the fitting region is shown in all the
figures in order to determine a direction of father approximation}$^)$.
\begin{table}[cbth]   % Table 1
\caption{Results of the experimental data fitting at $a_4,~a_5,~a_6,~a_7=0$.}
$$
\begin{array}{|c|c|c|c|c|} \hline
                               &       a_1        &      a_2      &      a_3       & \chi^2/NOF \\ \hline
\bar pp, P_{lab}\geq 100~GeV/c  & 0.443 \pm 0.004  & 2.50 \pm 0.29 & 0.430 \pm 0.026&   14/22    \\ \hline
\bar pp, P_{lab}\geq ~10~GeV/c  & 0.453 \pm 0.003  & 3.71 \pm 0.12 & 0.515 \pm 0.009&   37/55    \\ \hline
\bar pp, P_{lab}\geq ~~5~GeV/c  & 0.459 \pm 0.002  & 4.16 \pm 0.11 & 0.545 \pm 0.008&   57/63    \\ \hline
\end{array}
$$
$$
\begin{array}{|c|c|c|c|c|} \hline
                               &       a_1        &      a_2      &      a_3       & \chi^2/NOF \\ \hline
 pp, P_{lab}\geq 100~GeV/c  & 0.432 \pm 0.004  & 1.17 \pm 0.110& 0.356 \pm 0.024&   30/56    \\ \hline
 pp, P_{lab}\geq ~10~GeV/c  & 0.429 \pm 0.002  & 1.10 \pm 0.020& 0.341 \pm 0.006&   55/109   \\ \hline
 pp, P_{lab}\geq ~~5~GeV/c  & 0.424 \pm 0.001  & 1.04 \pm 0.005& 0.320 \pm 0.003&  111/125   \\ \hline
 \end{array}
$$
\end{table}

As seen, the parameters $a_1$ for $\bar pp$ and $pp$ interactions
become close to each other with the energy growth. This is not
observed for the parameters $a_2$ and $a_3$. To crown it all, the
small value of the parameter $a_3$ for the $pp$ interactions points
out that we need to increase the imaginal part of the phase function
taking additional exchanges. This contradicts with the starting
assumptions. The contradiction can be erased at fixing the parameter
$a_3$ at the level 0.5. Results of the corresponding fit are given in
table 2, and presented in Fig. 5 by the dashed lines. As seen,
the results allows one to hope on a father success.

Let us include the additional exchanges. The results of the fit of the
experimental data at $a_3=0.5$ and $a_5$, $a_6$, $a_7=0$ are presented
in table 3 and in Fig. 6 by solid lines. As seen, a significant rapprochement of
the pomeron exchange parameters has occurred. It is observed a
difference between the parameters $a_2$ as before. At the same time at
$p_{lab}\geq 5$ GeV/c the parameters of the trajectory responsible for
the process of Fig. 2c become close. It is clear that the next
additional exchange must have the positive signature.

\begin{table}[cbth]   % Table 2
\caption{Results of the experimental data fitting at $a_3=0.5$, $a_4,~a_5,~a_6,~a_7=0$.}
$$
\begin{array}{|c|c|c|c|c|} \hline
                               &       a_1        &      a_2      &      a_3       & \chi^2/NOF \\ \hline
\bar pp, P_{lab}\geq ~10~GeV/c & 0.449 \pm 0.001  & 3.54 \pm 0.022& 0.5            &   38/56    \\ \hline
 pp, P_{lab}\geq ~10~GeV/c     & 0.464 \pm 0.003  & 1.718\pm 0.006& 0.5            &  388/110   \\ \hline
 \end{array}
$$                    % Table 3
\caption{Results of the experimental data fitting at $a_3=0.5$, $a_5,~a_6,~a_7=0$.}
$$
\begin{array}{|c|c|c|c|c|c|} \hline
                                &       a_1        &      a_2      &      a_4       & \chi^2/NOF \\ \hline
\bar pp, P_{lab}\geq ~10~GeV/c  & 0.453 \pm 0.002  & 3.37 \pm 0.08 & 1.06 \pm 0.46 &   35/55    \\ \hline
\bar pp, P_{lab}\geq ~~5~GeV/c  & 0.456 \pm 0.002  & 3.19 \pm 0.06 & 2.29 \pm 0.32 &   48/62    \\ \hline
\bar pp, P_{lab}\geq ~~4~GeV/c  & 0.461 \pm 0.002  & 3.00 \pm 0.05 & 3.50 \pm 0.20 &   90/66    \\ \hline
\end{array}
$$
$$
\begin{array}{|c|c|c|c|c|c|} \hline
                           &       a_1        &      a_2      &      a_4       & \chi^2/NOF \\ \hline
pp, P_{lab}\geq ~10~GeV/c  & 0.450 \pm 0.001  & 2.34 \pm 0.03 &  3.27 \pm 0.13 &   73/109   \\ \hline
pp, P_{lab}\geq ~~5~GeV/c  & 0.454 \pm 0.001  & 2.16 \pm 0.01 &  2.33 \pm 0.04 &  178/125   \\ \hline
pp, P_{lab}\geq ~~4~GeV/c  & 0.457 \pm 0.001  & 2.06 \pm 0.01 &  1.93 \pm 0.04 &  344/133   \\ \hline
\end{array}
$$
\end{table}

Results of the fit with four terms of the phase function are
presented in table 4. As seen, the fit does not give a
self-consistent set of the parameters for $\bar pp$ and $pp$
interactions.
Summing up we choose at $p_{lab}\geq 5$ GeV/c the set of the
parameters presented in table 5.
\begin{figure}[cbth] % Fig. 5
\begin{center}
\psfig{file=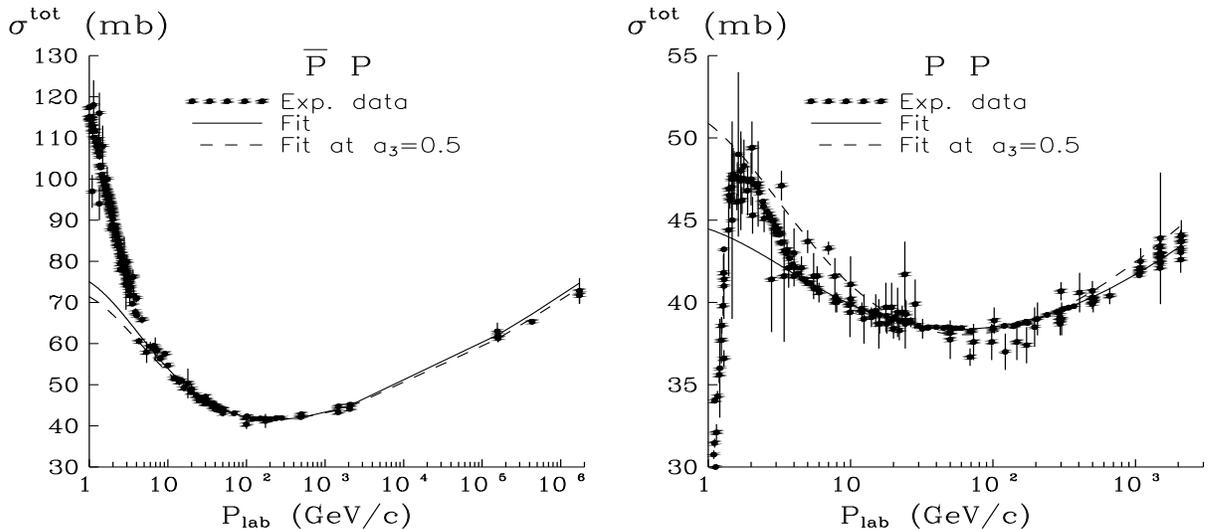,width=160mm,height=70mm,angle=0}
\caption{Results of the fitting with two first terms of the phase function.}
\end{center}
\end{figure}

Let us consider a pessimistic variant with $\alpha_a=-0.5$. Doing the
same as it was before, we have results presented in table 6 and shown
in Fig. 6 by the dashed lines. Once more, as it was before the
parameters $a_1$ and $a_2$ become close at $p_{lab}\sim 5$ GeV/c. In
the same manner we determine that the additional exchange must have
the positive signature. Though taking into account the fifth term of
the phase function does not lead to a rapprochement of the parameters
for the $\bar pp$ and $pp$ interactions. Therefore we propose for
the region of $p_{lab}\geq 5$ GeV/c the parameter set presented in
table 7 for the considered case.

\begin{table}[cbth] % Табл. 4
\caption{Results of the experimental data fitting at $a_3=0.5$, $a_6,~a_7=0$.}
$$
\begin{array}{|c|c|c|c|c|c|} \hline
                                &       a_1        &      a_2      &      a_4      &       a_5      & \chi^2/NOF \\ \hline
\bar pp, P_{lab}\geq ~10~GeV/c  & 0.449 \pm 0.002  & 3.65 \pm 0.16 &-2.66 \pm 1.87 & 12.8  \pm 6.1  &   34/54    \\ \hline
\bar pp, P_{lab}\geq ~~5~GeV/c  & 0.450 \pm 0.002  & 3.63 \pm 0.13 &-2.57 \pm 1.26 & 13.2  \pm 3.2  &   41/61    \\ \hline
\bar pp, P_{lab}\geq ~~4~GeV/c  & 0.450 \pm 0.002  & 3.61 \pm 0.10 &-2.27 \pm 0.89 & 12.2  \pm 1.72 &   71/65    \\ \hline
\bar pp, P_{lab}\geq ~~3~GeV/c  & 0.435 \pm 0.002  & 4.48 \pm 0.07 &-10.5 \pm 0.5  & 28.1  \pm 0.6  &  173/80    \\ \hline
\end{array}
$$
$$
\begin{array}{|c|c|c|c|c|c|} \hline
                           &       a_1        &      a_2      &      a_4       &       a_5       &\chi^2/NOF \\ \hline
pp, P_{lab}\geq ~10~GeV/c  & 0.445 \pm 0.001  & 2.67 \pm 0.07 &  7.11 \pm 0.74 &  11.59 \pm 2.14 &  60/108   \\ \hline
pp, P_{lab}\geq ~~5~GeV/c  & 0.446 \pm 0.001  & 2.58 \pm 0.04 &  5.92 \pm 0.28 &  ~7.37 \pm 0.54 & 105/124   \\ \hline
pp, P_{lab}\geq ~~4~GeV/c  & 0.445 \pm 0.001  & 2.62 \pm 0.03 &  6.24 \pm 0.19 &  ~7.97 \pm 0.32 & 119/132   \\ \hline
pp, P_{lab}\geq ~~3~GeV/c  & 0.440 \pm 0.001  & 2.88 \pm 0.02 &  8.24 \pm 0.10 &  11.57 \pm 0.13 & 253/147   \\ \hline
\end{array}
$$
%\end{table}
%\begin{table}[cbth]    % Табл. 5
\caption{Set of proposed parameters No 1.}
$$
\begin{array}{|c|c|c|c|c|c|c|c|} \hline
                                &  a_1  & a_2  & a_3 & a_4  & a_5  & a_6  & a_7  \\ \hline
\bar pp, P_{lab}\geq ~~5~GeV/c  & 0.455 & 3.19 & 0.5 & 2.31 & 0.   & 0.   & 0.   \\ \hline
pp, P_{lab}\geq ~~5~GeV/c       & 0.455 & 2.16 & 0.5 & 2.31 & 0.   & 0.   & 0.   \\ \hline
\end{array}
$$
\end{table}

\begin{figure}[b] %[cbth]
\begin{center}
\psfig{file=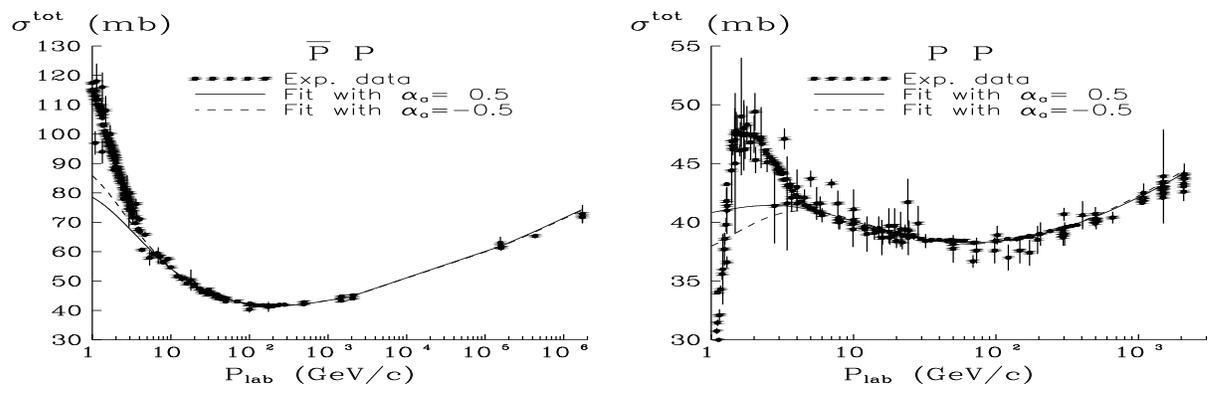,width=160mm,height=50mm,angle=0}
\caption{Results of the fitting with three terms of the phase function.}
\end{center}
\end{figure}
\newpage

\begin{table}[cbth]                       % Табл. 6
\caption{Results of the experimental data fitting at $a_3=0.5$, $a_4,~a_6,~a_7=0$.}
$$
\begin{array}{|c|c|c|c|c|c|} \hline
                                &       a_1        &      a_2      &      a_5       & \chi^2/NOF \\ \hline
\bar pp, P_{lab}\geq ~10~GeV/c  & 0.452 \pm 0.002  & 3.44 \pm 0.04 & 4.33 \pm 1.64 &   34/55    \\ \hline
\bar pp, P_{lab}\geq ~~5~GeV/c  & 0.454 \pm 0.002  & 3.39 \pm 0.03 & 7.12 \pm 0.89 &   43/62    \\ \hline
\bar pp, P_{lab}\geq ~~4~GeV/c  & 0.455 \pm 0.001  & 3.36 \pm 0.03 & 8.11 \pm 0.41 &   74/66    \\ \hline
\end{array}
$$
$$
\begin{array}{|c|c|c|c|c|c|} \hline
                           &       a_1        &      a_2      &      a_5       & \chi^2/NOF \\ \hline
pp, P_{lab}\geq ~10~GeV/c  & 0.454 \pm 0.001  & 2.04 \pm 0.02 & 10.4  \pm 0.4  &  100/109   \\ \hline
pp, P_{lab}\geq ~~5~GeV/c  & 0.459 \pm 0.001  & 1.88 \pm 0.01 &  5.29 \pm 0.11 &  279/125   \\ \hline
pp, P_{lab}\geq ~~4~GeV/c  & 0.463 \pm 0.001  & 1.79 \pm 0.01 &  3.91 \pm 0.08 &  530/133   \\ \hline
\end{array}
$$
\caption{Set of proposed parameters No 2.} % Table 7
$$
\begin{array}{|c|c|c|c|c|c|c|c|} \hline
                                &  a_1  & a_2  & a_3 & a_4  & a_5  & a_6  & a_7  \\ \hline
\bar pp, P_{lab}\geq ~~5~GeV/c  & 0.457 & 3.39 & 0.5 & 0.   & 6.2  & 0.   & 0.   \\ \hline
pp, P_{lab}\geq ~~5~GeV/c       & 0.457 & 1.88 & 0.5 & 0.   & 6.2  & 0.   & 0.   \\ \hline
\end{array}
$$
\end{table}

An explanation is needed for the difference of the parameters $a_2$
for $\bar pp$ and $pp$ interaction. Usually it is assumed that the
contribution of the one-meson exchange consists from yields of the two
poles with negative and positive signatures:
$F_{\bar pp} = F_P + F_f + F_\omega$,
$F_{pp}      = F_P + F_f - F_\omega$.
In accordance with this one can introduce
$a_{2f}=(a_{2\bar pp}+a_{2pp})/2$ and $a_{2\omega}=(a_{2\bar
pp}-a_{2pp})/2$ and repeat the fit once more. However this procedure
is rather unstable. It is needed to look at other set of experimental
data what is out of the scope of our paper. The simplest
receipt for overcoming the problem is neglection by the real part of the
second term of the phase function. An apparently consequence of the
assumption is that the ratios of the real to imaginary parts of the
elastic scattering amplitude for $\bar pp$ and $pp$ interactions will
be near to zero in the studied energy range. At the same time, it is well
known from experiment that the ratio for the $\bar pp$ interactions has
a positive value at low energies, but the ratio for the $pp$ interactions
has a negative value.

The more complicate situation takes place with a description of the
elastic scattering cross section (see Fig. 7), and with a description
of the data at $p_{lab}\le 5$ GeV/c. Here one needs to assume
various interaction radii of different reggeons with nucleon.

If we restrict ourself by the region of $p_{lab}\geq 5$ GeV/c, we
believe that the reached results are acceptable. The results are shown
in Fig. 7.

\section{Cross sections of $\bar pp$ interactions}
Cross sections of the processes of Fig. 2b, 2c and 2g obtained with
a help of Eq. (13) are presented in Fig. 8. There are also inelastic
cross sections of $\bar pp$ interactions calculated with the HERA-CERN
parametrization \cite{PDG_param} (upper solid curve) and with the
parameter set No 1 (upper dashed curve). As seen, the two parametrization
are close to each other.

Let us note that the screening corrections (Eq. (13)) change the
energy dependencies of the processes of Fig. 2b and 2c. For example,
the cross section of the process of Fig. 2b falls down more slowly
than $1/\sqrt{s}$ with energy increase. One needs to take these into
account at a calculation of cross sections of other reactions, e.a.
$\bar pp \rightarrow K^+ K^-$, $\bar pp \rightarrow  \bar \Lambda
\Lambda$, in the spirit of Ref. \cite{Kaidalov_Volkov}.
\begin{figure}[cbth]
\begin{center}
\psfig{file=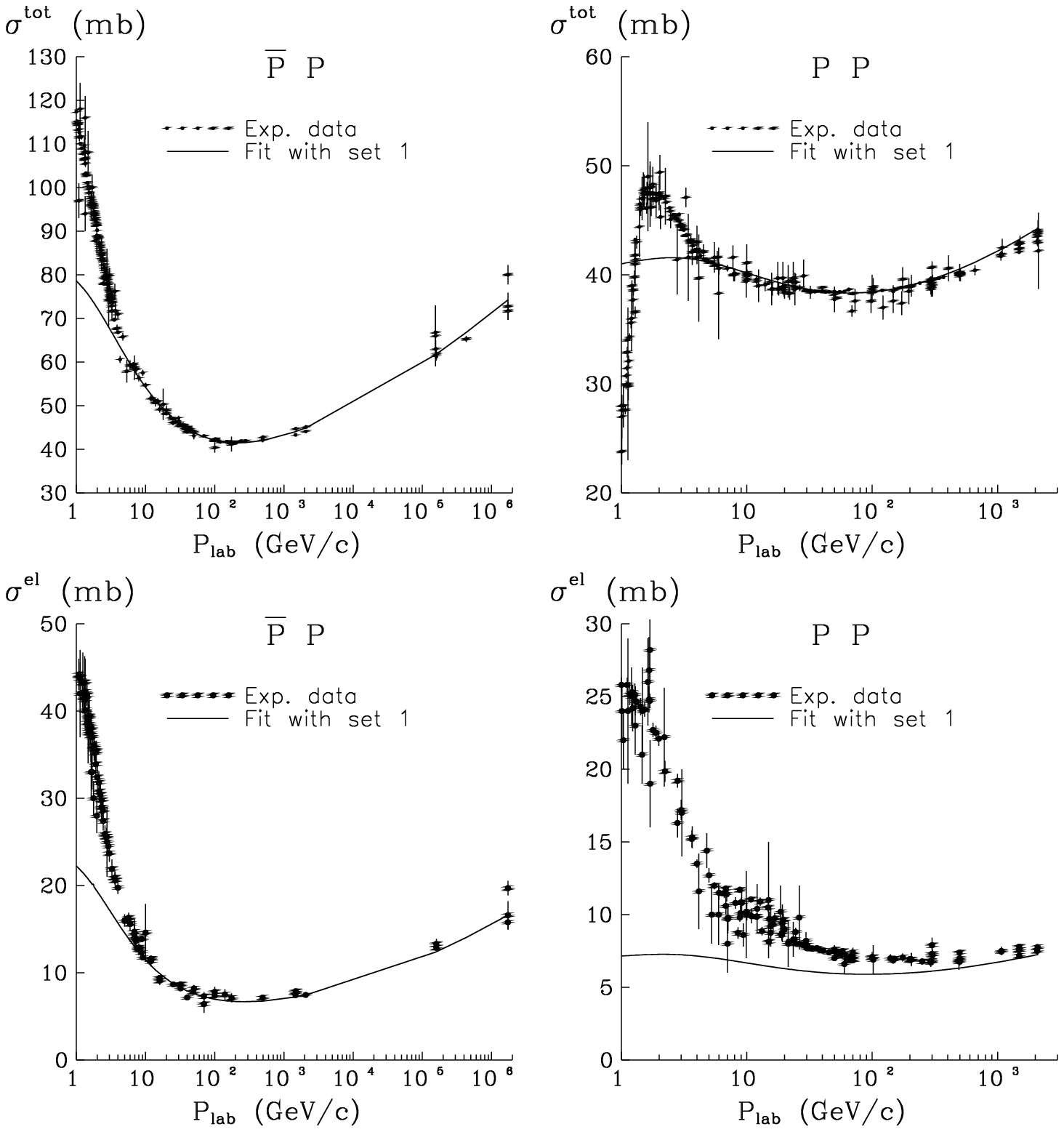,width=160mm,height=160mm,angle=0}
\caption{Final results of the experimental data fitting with the parameter set No 1.}
\end{center}
\end{figure}

It is a subject of specific interest to describe annihilation cross
section, the cross section of reaction without baryons in the final
state. The known experimental data \cite{Flaminio} are presented in
Fig. 8 by the open points. The parametrization of the form $1/s$ used
in the Ultrarelativistic Quantum Molecular Dynamic (UrQMD) model is
given by filled points. As seen, the annihilation can not be explained by
the process of Fig. 2c. It can not be a constant part of the processes
connected with the second term of the phase function due to different
energy dependencies of the annihilation and the cross section of the
process of Fig. 2b.
It can not be a part of the pomeron exchange, as
it is often assumed starting from Ref. \cite{Eylon}.
One needs to
assume a new mechanism of the baryon annihilation.
\begin{figure}[t] %[cbth]
\begin{center}
\psfig{file=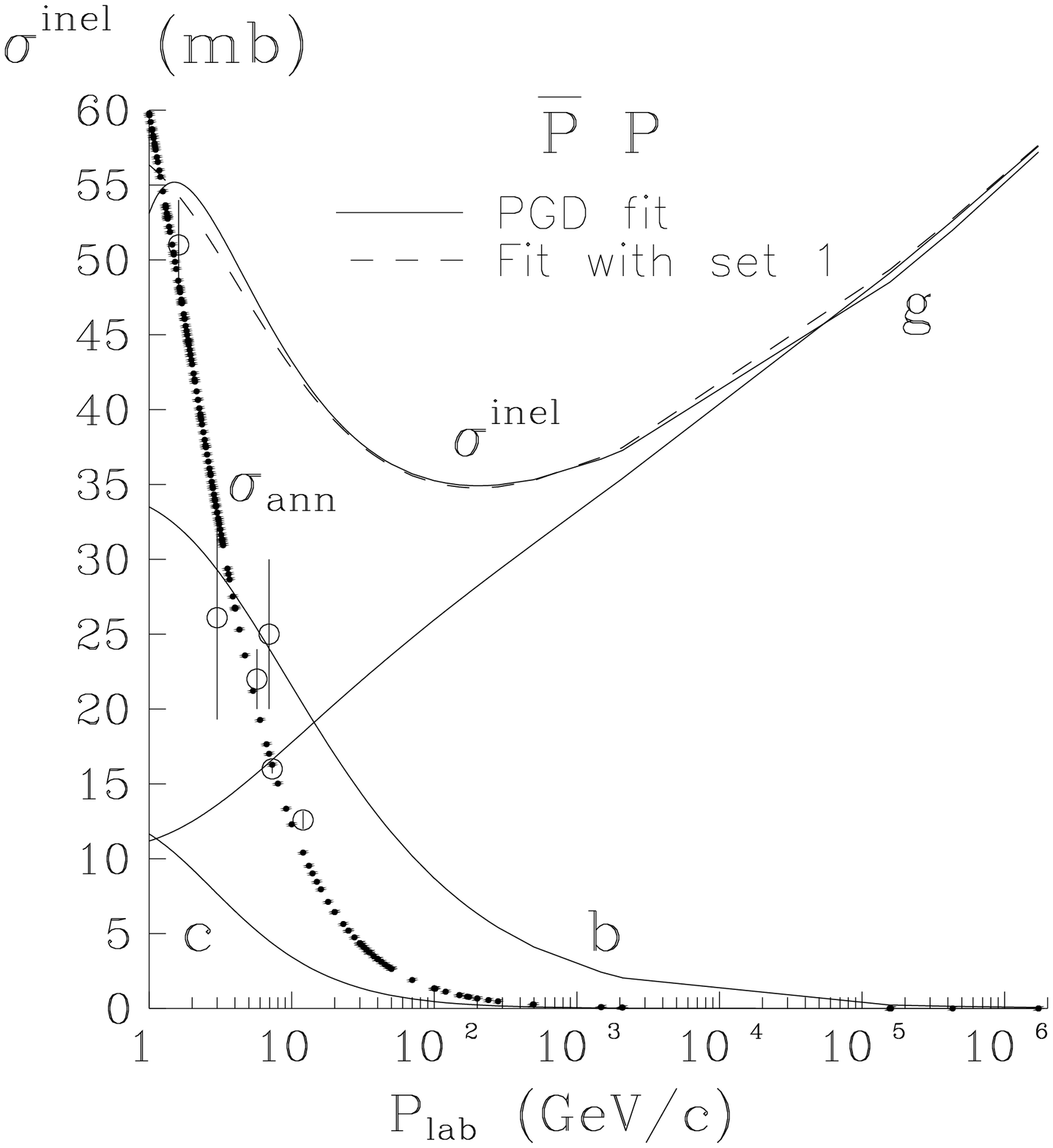,width=160mm,height=130mm,angle=0}
\caption{Decomposition of the inelastic $\bar pp$-interaction cross section.
Curves b, c, g represent cross sections of the processes 2b, 2c, 2g. Filled
points are UrQMD model parametrization \protect{\cite{URQMD}} of the
annihilation cross section. Open points are the experimental data
\protect{\cite{Flaminio}}.}
\end{center}
\end{figure}

In Refs. \cite{Kopeliovich,Echaya} it was assumed that at intermediate
energies the main contribution to the annihilation cross section is
given be the process with quark and antiquark annihilation with
following one-gluon exchange (see Fig. 9a). However the gluon exchange
can not change the energy dependence of the type $1/\sqrt{s}$
\cite{Kopeliovich,Echaya}.

We believe that diquark or anti-diquark can emit a meson, and after
that the string junction annihilation takes place (see Fig. 9b).
Nearly the same mechanism is accepted in Ur.QMD model \cite{URQMD}
which describes quite well experimental data at intermediate energies.
Because diquark -- anti-diquarks strings are created in the processes
of Fig. 2b and 2g, the proposed mechanism can be connected with the
main processes of $\bar pp$ interactions. The energy dependence of the
annihilation can be determined by the fragmentation of the strings. We
are going to consider the fragmentation in following paper.
\begin{figure}[cbth]
\begin{center}
\psfig{file=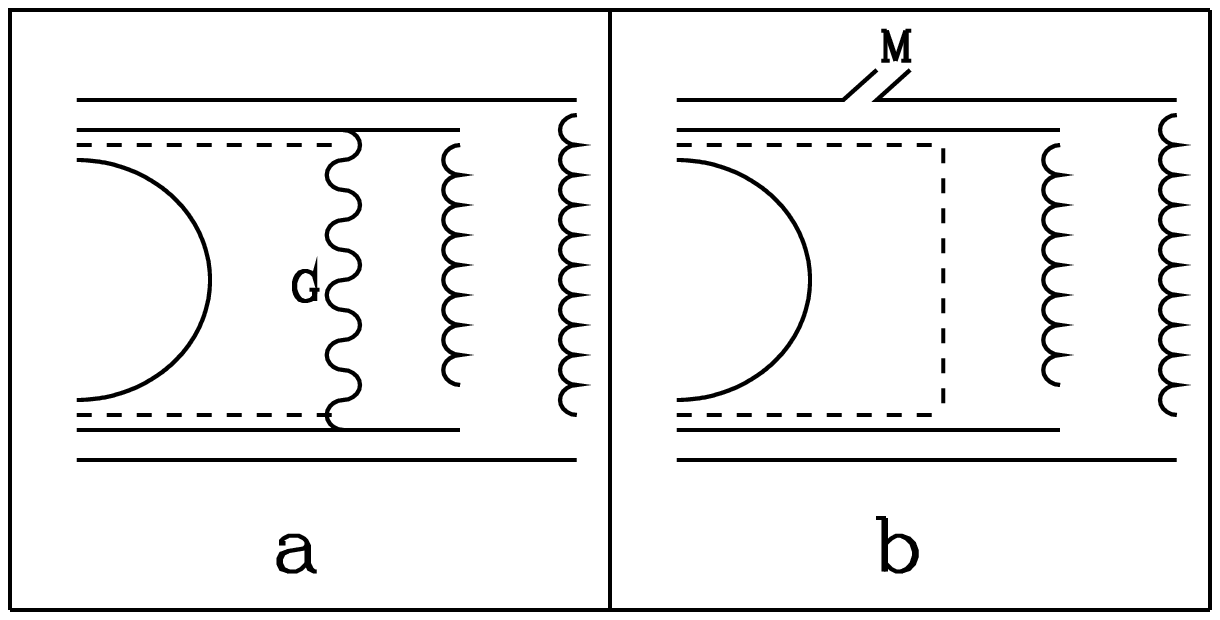,width=160mm,height=50mm,angle=0}
\caption{Possible mechanisms of baryon annihilation.}
\end{center}
\end{figure}

\section*{Summary}
\begin{itemize}
\item
The expression is proposed for the calculation of the cross sections
of various reactions in $\bar pp$ interactions, for the processes
with creation of one or two strings, within the framework of the
eikonal approach in the assumption of the identical interaction radii
of different reggeon with nucleon. The cross sections are needed for
Monte Carlo simulation of the reactions.

\item
The parameters of the pomeron exchange, meson exchange, and exchange
of a meson with hidden baryon number are determined at fitting the
experimental data on the total $\bar pp$ and $pp$ interaction cross
sections.

\item
Using the fitted parameters, the cross sections of $\bar pp$ interactions
are calculated at $p_{lab}\geq 5$ GeV/c.

\end{itemize}

The authors are thankful to E. Strokovsky, J. Ritman and M. Sapozhnikov
for stimulating discussions, and A.B. Kaidalov and K.G. Boreskov for
consideration of the paper. We thank A. Polanski for
his support and interest to the paper.

\end{document}